\documentclass[aps,preprint,showkeys,nofootinbib,superscriptaddress,amsmath,amssymb]{revtex4-2}

\usepackage[utf8]{inputenc}
\usepackage[T1]{fontenc}
\usepackage{amsmath,amssymb,amsfonts,bm}
\usepackage{graphicx}
\graphicspath{{figures/}}
\usepackage{lmodern}
\usepackage{microtype}
\usepackage{xcolor}
\usepackage{hyperref}
\usepackage{booktabs}
\usepackage{array}

\hypersetup{colorlinks=true,linkcolor=blue,citecolor=blue,urlcolor=blue,
 pdftitle={Klein-Gordon and Dirac Oscillators in a Constant-Curvature Fock-Lorentz Background},
 pdfauthor={A. Boumali, N. Jafari, M. Botshekananfard}}

\newcommand{\be}{\begin{equation}}
\newcommand{\ee}{\end{equation}}
\newcommand{\bea}{\begin{eqnarray}}
\newcommand{\eea}{\end{eqnarray}}
\newcommand{\vp}{\mathbf{p}}
\newcommand{\vx}{\mathbf{x}}
\newcommand{\vy}{\mathbf{y}}
\newcommand{\vk}{\mathbf{k}}
\newcommand{\mapp}{m_{\mathrm{app}}}
\newcommand{\Ric}{\mathcal{R}}
\newcommand{\Ham}{\mathcal{H}}
\newcommand{\dd}{\mathrm{d}}

\begin{document}

\title{Klein--Gordon and Dirac Oscillators in a Constant-Curvature Fock--Lorentz Background}

\author{A. Boumali}
\email{abdelmalek.boumali@univ-tebessa.dz}
\thanks{ORCID: \href{https://orcid.org/0000-0003-2552-0427}{0000-0003-2552-0427}}
\affiliation{Laboratory of Applied Physics and Theoretical Physics, Department of Matter Sciences, Faculty of Exact Sciences and Sciences of Nature and Life, Larbi Tebessi University, Tebessa, Algeria}

\author{N. Jafari}
\email{nosrat.jafari@fai.kz}
\thanks{ORCID: \href{https://orcid.org/0000-0002-0285-6107}{0000-0002-0285-6107}}
\affiliation{Fesenkov Astrophysical Institute, 050020 Almaty, Kazakhstan}
\affiliation{Al-Farabi Kazakh National University, Al-Farabi Avenue 71, 050040 Almaty, Kazakhstan}

\author{Manizheh Botshekananfard}
\email{manizheh.botshekananfard@bogazici.edu.tr}
\thanks{ORCID: \href{https://orcid.org/0000-0003-1676-6070}{0000-0003-1676-6070}}
\affiliation{Department of Physics, Bo\u{g}azi\c{c}i University, Bebek, 34342, \.{I}stanbul, Turkey}

\date{\today}

\begin{abstract}
Klein--Gordon and Dirac oscillators are formulated on the constant-curvature spacetime $g_{\mu\nu}=s^{-2}(t)\eta_{\mu\nu}$ generated by the Fock--Lorentz conformal factor $s(t)=1+ct/R$, a conformal Poincar\'e patch of de Sitter spacetime. The metric is a geometric postulate rather than a consequence of the Fock--Lorentz transformations, whose projective map pulls Minkowski space back to a flat metric. With the ordinary cotangent-bundle phase space, $(E^{2}-\vp^{2}c^{2})s^{2}=m_{0}^{2}c^{4}$ is the conformal-coordinate form of the covariant mass shell and not a deformed dispersion relation. Covariant actions give the canonically rescaled scalar equation, with its order-$R^{-2}$ curvature term, and the Dirac equation with $\mapp(t)=m_{0}/s(t)$. The oscillator couplings are introduced in the conformal chart and define coordinate-dependent simplified models; their fixed-time spectra follow in closed form, and the Klein--Gordon sector is treated as a second-order mode system with adiabaticity assessed using the level and particle--antiparticle gaps. The spectral drift at fixed coordinate frequency is removed by a change of prescription together with a rescaling, so the rescaled eigenvalue is not a locally measured spectrum.
\end{abstract}

\keywords{Fock--Lorentz coordinates; de Sitter spacetime; curved-spacetime quantum field theory; Klein--Gordon oscillator; Dirac oscillator; spin connection; coordinate-defined oscillator models; adiabatic dynamics.}

\maketitle

\section{Introduction}
\label{sec:intro}

The Fock--Lorentz (FL) transformations provide a fractional-linear form of relativistic coordinate transformations that supplements the invariant speed of light with a second, large invariant length scale $R$~\cite{Fock,Manida,Stepanov}. In the time-like sector their characteristic dependence is governed by the conformal factor
\be
 s(t)=1+\frac{ct}{R},
\label{eq:s-intro}
\ee
so that invariant kinematic quantities often involve the combination $(c^{2}t^{2}-\vx^{2})/s^{2}(t)$. This structure has motivated studies on relativistic wave equations, time-dependent effective masses, and exactly solvable oscillator systems on FL-type backgrounds. Its quantum-mechanical formulation is not unique: whether $s(t)$ is treated as a coordinate relabelling of Minkowski space, as a genuine curved geometry, or as a deformation of the momentum sector leads to physically inequivalent theories. A consistent quantum treatment therefore requires that the spacetime geometry, the phase space, the field variables, and the class of observable quantities be clearly specified.

Several distinct frameworks describe such nonlinear relativistic transformations. They differ in where the nonlinearity enters. Doubly special relativity deforms the momentum sector; gravity's-rainbow models promote the frame fields, and hence the metric, to energy-dependent objects; and relative-locality models introduce the nonlinearity through a curved momentum space with momentum-dependent localization~\cite{Rainbow,RelLoc}. Nonlinear position-space realizations are frequently implemented with a deformed invariant pairing between coordinates and momenta~\cite{MSPosition}. Such a deformation applies to particular model classes and is not required by nonlinearity alone: minimum-length constructions exist in which position, time, energy and momentum operators are modified together while both the canonical commutator and the standard dispersion relation are retained~\cite{Bishop2021,Bishop2022}. Statements below concerning deformed symplectic structures, noncanonical pairings or modified dispersion relations apply only to those model classes. These frameworks modify the canonical phase-space structure, making it harder to distinguish physical effects in an FL oscillator from effects of the chosen deformation. In this work, the FL conformal factor is used to define a \emph{single, energy-independent} spacetime metric,
\be
 g_{\mu\nu}(x)=a^{2}(t)\eta_{\mu\nu},\qquad
 a(t)=\frac{1}{s(t)},\qquad s(t)=1+\frac{ct}{R},
\label{eq:metric-intro}
\ee
with $\eta_{\mu\nu}=\mathrm{diag}(1,-1,-1,-1)$ and $s(t)>0$. This metric has constant curvature and is locally equivalent to the spatially flat conformal (Poincar\'e) chart of four-dimensional de Sitter spacetime~\cite{Foughali,Pascu}. Its phase space is the ordinary cotangent bundle $T^{*}{\cal M}$ with canonical brackets, and the scalar and spinor dynamics follow from conventional curved-spacetime actions~\cite{Birrell,ParkerToms,WaldQFT}. No deformation of the symplectic structure or of the position--momentum pairing is introduced anywhere in the construction.

Equation~(\ref{eq:metric-intro}) is an independent geometric assumption within the FL framework. The metric retains the conformal part of the FL structure and uses it to define a spacetime geometry. Treating the projective FL map $X^{\mu}=x^{\mu}/s$ as a coordinate transformation of Minkowski space produces the flat pullback obtained in Sec.~\ref{sec:geometry}; retaining only its conformal part is a separate modelling step, and the same conformal factor admits other completions. Whether the FL transformations act as isometries of Eq.~(\ref{eq:metric-intro}) is not determined here, and no result below depends on it.

Two consequences follow. First, Secs.~\ref{sec:geometry}--\ref{sec:Dirac} are quantum field theory on a de Sitter conformal chart: the curvature tensor, the Hubble-type term, the canonical rescalings and the spin connection are standard results, recorded in the FL parametrization because the later sections require them in that form. Second, the phrase ``constant-curvature Fock--Lorentz background'' denotes exactly this object---the constant-curvature metric built from the FL conformal factor---and carries no implication that FL kinematics implies the curvature.

Three aspects of the analysis are specific to FL. (i) The chart: the scale factor is the FL factor $a=1/s$ and the conformal time is FL coordinate time, so the quantities computed below are expressed directly in FL variables. (ii) The coefficient $\mapp(t)=m_{0}/s(t)$ generated by this chart in the canonically rescaled field equations, usually called the FL ``apparent mass'', is shown below to be a conformal-coordinate parameter rather than a varying invariant rest mass. (iii) The operational question raised in the FL literature: what is measured by a trap whose frequency is held fixed with respect to FL coordinate time, compared with one held fixed in the local frame. The answer obtained in Sec.~\ref{sec:operational} is that the spectral drift produced by the first prescription follows from the chart and from the prescription.

The geometric formulation clarifies the meaning of the effective mass. The conformal geometry naturally generates the coefficient
\be
 \mapp(t)=m_{0}a(t)=\frac{m_{0}}{1+ct/R},
\label{eq:mapp-intro}
\ee
in the canonically rescaled field equations. We stress that $\mapp(t)$ is a conformal-coordinate mass parameter and \emph{not} a time-dependent invariant rest mass. In the same coordinates the point-particle mass shell reads
\be
 (E^{2}-\vp^{2}c^{2})s^{2}(t)=m_{0}^{2}c^{4},
\label{eq:massshell-intro}
\ee
which is the coordinate representation of the manifestly covariant relation $g^{\mu\nu}p_{\mu}p_{\nu}=m_{0}^{2}c^{2}$; projection onto a local orthonormal tetrad restores the standard special-relativistic dispersion relation with the constant rest mass $m_{0}$. The time dependence in Eqs.~(\ref{eq:mapp-intro}) and~(\ref{eq:massshell-intro}) therefore comes entirely from the conformal chart. This distinction guides the interpretation of the quantities computed below.

The geometry enters the scalar and spinor equations through related mechanisms. The covariant Klein--Gordon equation acquires a first-derivative term proportional to the conformal Hubble function; a canonical field rescaling eliminates this term while leaving behind a definite curvature contribution of order $R^{-2}$. For the Dirac field the corresponding geometric information enters through the spin connection: the standard conformal spinor rescaling produces a first-order equation of flat form with the coefficient $\mapp(t)$, and squaring the operator generates a correction proportional to $\dot{\mapp}\gamma^{0}$. Identifying a common geometric origin for the time-dependent terms in the two relativistic wave equations is one of the results of the analysis.

The Klein--Gordon and Dirac oscillator couplings are introduced as exactly solvable models on this background. Because both models remain exactly solvable at fixed time, they permit a simultaneous treatment of instantaneous spectra, localization lengths, adiabatic transport, and the operational distinction between coordinate-time and proper-time laboratory protocols. A frequency held fixed with respect to conformal coordinate time and a frequency held fixed in the local orthonormal frame describe physically distinct trapping devices; comparing them isolates the spectral variations that are mere consequences of the conformal chart from those that persist as invariant curvature or tidal effects. This operational separation is the physical question that motivates the entire construction.

The aim of this work is to determine the spectra of these coordinate-defined models and clarify their interpretation. We construct and solve in closed form, at fixed time, Klein--Gordon and Dirac oscillator models \emph{defined in the FL conformal chart}, and determine what their eigenvalues represent. Establishing which spectral properties of a physical relativistic oscillator in de Sitter spacetime are invariant is a distinct problem, requiring a covariantly specified trapping field, a clock and a detector; the required ingredients are described in Secs.~\ref{sec:osc-status} and~\ref{sec:limits}. The result obtained here is that the spectral drift produced by holding the conformal-coordinate trap frequency fixed follows from the chart and from the chosen frequency prescription, and we specify the additional ingredients required to connect this result to measurements.

The analysis has four main results. First, we record the constant-curvature geometry generated by the FL conformal factor, make its de Sitter interpretation explicit, and distinguish it from the flat pullback obtained by treating $s(t)$ as a coordinate map. Second, we obtain the covariant Klein--Gordon and Dirac equations from action functionals, identifying the Hubble-type and spin-connection terms and their removal by canonical field redefinitions, and we identify the origin of each remaining term to clarify its interpretation. Third, we obtain closed-form fixed-time spectra, eigenfunctions, localization lengths and adiabaticity conditions for both oscillator sectors, retaining the complete pair of Klein--Gordon factor orderings and both Dirac branches with the isolated level, and treating the Klein--Gordon equation as a second-order system. Fourth, we compare fixed coordinate-frequency and fixed local-frequency prescriptions, show that the leading drift is removed from the rescaled eigenvalue, and clarify the meaning of the remaining quantity. A structural comparison with doubly special relativity, gravity's rainbow, nonlinear position relativity and relative locality places the model in context, with the qualification stated above for minimum-length models.

The remainder of the paper is organized as follows. Section~\ref{sec:geometry} develops the constant-curvature geometry and the canonical phase space. Sections~\ref{sec:KG} and~\ref{sec:Dirac} construct the covariant scalar and spinor field theories. Sections~\ref{sec:KGosc} and~\ref{sec:Diracosc} analyse the Klein--Gordon and Dirac oscillator sectors. Section~\ref{sec:operational} develops the operational observables and the protocol comparison, Sec.~\ref{sec:comparison} compares the model with nonlinear-relativity frameworks, and Sec.~\ref{sec:limits} assesses the relevant physical scales and the limitations of the analysis. The principal conclusions are summarized in Sec.~\ref{sec:conclusion}.

\section{Constant-curvature FL geometry and phase space}
\label{sec:geometry}

\subsection{Constant-curvature metric and de Sitter interpretation}

We use coordinates $x^{\mu}=(x^{0},\vx)$ with $x^{0}=ct$ and define
\be
 s(x^{0})=1+\frac{x^{0}}{R},\qquad a(x^{0})=s^{-1}(x^{0}).
\label{eq:sa}
\ee
The line element is
\be
 \dd\ell^{2}=g_{\mu\nu}\dd x^{\mu}\dd x^{\nu}
 =a^{2}(x^{0})\left[(\dd x^{0})^{2}-\dd\vx^{2}\right].
\label{eq:lineelement}
\ee
The nonvanishing connection coefficients may be written in terms of the conformal Hubble function $\Ham=a'/a=-1/(Rs)$, where a prime denotes $\dd/\dd x^{0}$:
\be
 \Gamma^{0}_{00}=\Ham,\qquad
 \Gamma^{0}_{ij}=\Ham\delta_{ij},\qquad
 \Gamma^{i}_{0j}=\Gamma^{i}_{j0}=\Ham\delta^{i}_{j}.
\label{eq:christoffel}
\ee
We define the Riemann tensor by
\be
 R^{\rho}{}_{\sigma\mu\nu}
 =\partial_{\mu}\Gamma^{\rho}_{\nu\sigma}
 -\partial_{\nu}\Gamma^{\rho}_{\mu\sigma}
 +\Gamma^{\rho}_{\mu\lambda}\Gamma^{\lambda}_{\nu\sigma}
 -\Gamma^{\rho}_{\nu\lambda}\Gamma^{\lambda}_{\mu\sigma}.
\label{eq:riemann-convention}
\ee
Direct calculation then gives
\bea
 R_{\mu\nu\rho\sigma}&=&-\frac{1}{R^{2}}
 (g_{\mu\rho}g_{\nu\sigma}-g_{\mu\sigma}g_{\nu\rho}),
\label{eq:riemann}\\
 R_{\mu\nu}&=&-\frac{3}{R^{2}}g_{\mu\nu},\qquad
 \Ric=-\frac{12}{R^{2}}.
\label{eq:ricci}
\eea
The sign of $\Ric$ reverses if the opposite Riemann-tensor convention is adopted; the invariant statement is that the sectional curvature has constant magnitude $R^{-2}$. To display the de Sitter character explicitly, introduce the shifted conformal time $\eta=t+R/c$. Equation~(\ref{eq:lineelement}) then becomes
\be
 \dd\ell^{2}=\frac{R^{2}}{c^{2}\eta^{2}}
 \left(c^{2}\dd\eta^{2}-\dd\vx^{2}\right),
\label{eq:dS-conformal-chart}
\ee
which is the spatially flat conformal (Poincar\'e) chart of de Sitter spacetime with curvature radius $R$. Reversing the orientation of $\eta$ maps the contracting representation adopted here onto the more familiar expanding chart. The surface $s=0$ is a conformal boundary of the chart rather than a genuine curvature singularity, since all curvature invariants remain finite there.

\begin{figure}[t]
\centering
\includegraphics[width=0.76\linewidth]{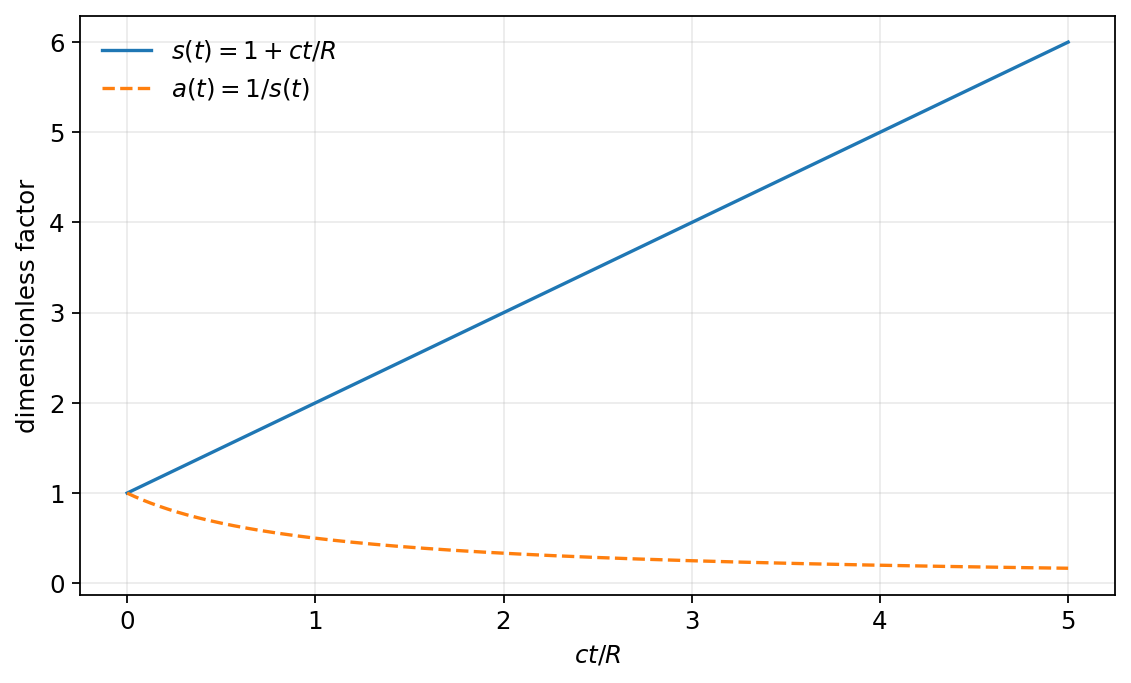}
\caption{The Fock--Lorentz conformal factor $s(t)=1+ct/R$ and the associated scale factor $a(t)=1/s(t)$ as functions of $ct/R$. The monotonic decrease of $a(t)$ characterizes the contracting orientation of the conformal chart used here; the value $s=0$ marks the conformal boundary of the patch.}
\label{fig:conformal-factor}
\end{figure}

\begin{figure}[t]
\centering
\includegraphics[width=0.78\linewidth]{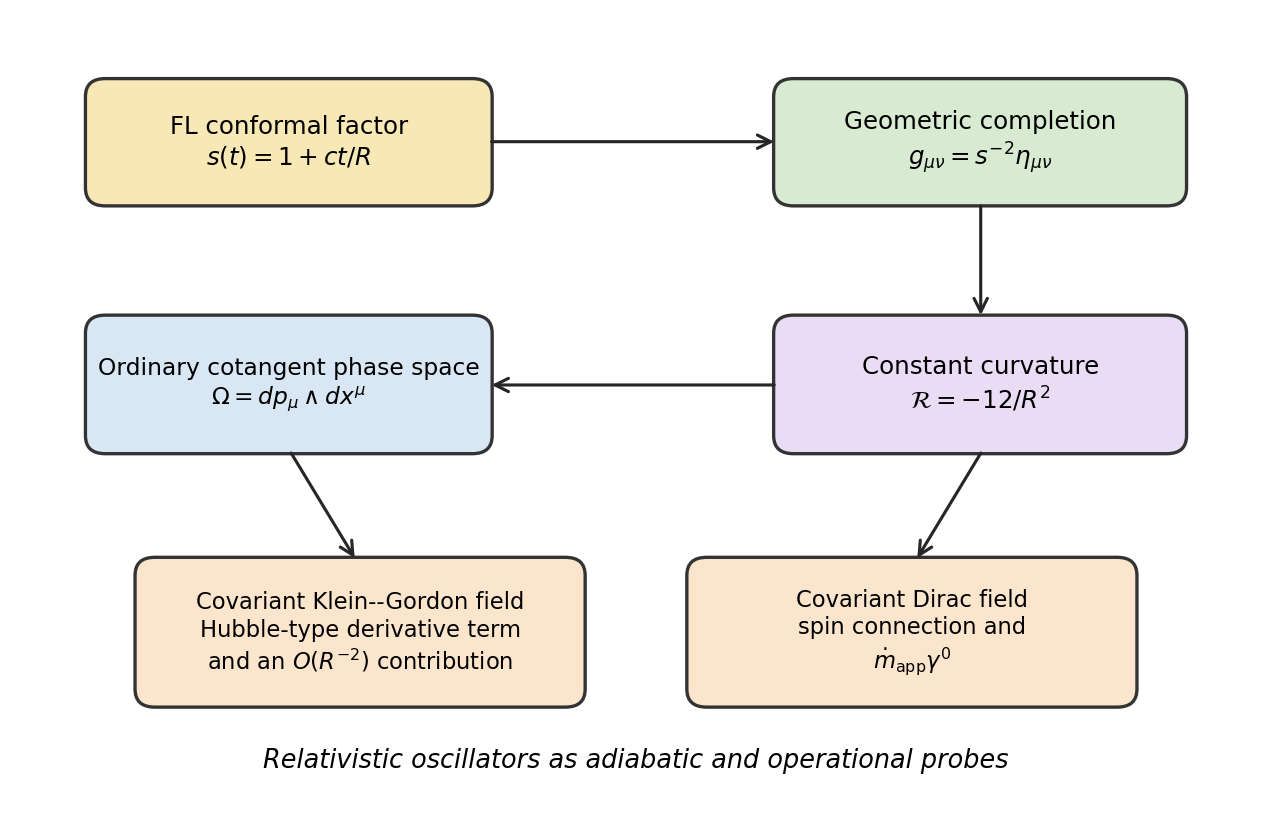}
\caption{Geometric structure adopted in this work. The FL factor is promoted to a conformal spacetime metric with constant curvature. The phase space is the ordinary cotangent bundle, and the scalar and spinor equations follow from covariant actions. No formal interchange of position and momentum variables is used.}
\label{fig:framework}
\end{figure}

\subsection{Geometric completion and projective pullback}

The metric in Eq.~(\ref{eq:lineelement}) must be distinguished from the metric obtained by treating
\be
 X^{\mu}=\frac{x^{\mu}}{s(x^{0})}
\label{eq:projective-map}
\ee
as a coordinate transformation of flat Minkowski spacetime. The pullback of $\eta_{\alpha\beta}\dd X^{\alpha}\dd X^{\beta}$ is
\be
 g^{\rm pull}_{\mu\nu}
 =\frac{\eta_{\mu\nu}}{s^{2}}
 -\frac{\delta^{0}_{\mu}x_{\nu}+\delta^{0}_{\nu}x_{\mu}}{Rs^{3}}
 +\frac{x^{2}}{R^{2}s^{4}}\delta^{0}_{\mu}\delta^{0}_{\nu}.
\label{eq:pullback}
\ee
Because Eq.~(\ref{eq:pullback}) is a coordinate pullback of Minkowski space, its curvature vanishes wherever the map is regular. By contrast, Eq.~(\ref{eq:lineelement}) retains only the conformal part and has the nonzero curvature in Eq.~(\ref{eq:ricci}). The two constructions are physically inequivalent. The analysis is restricted to the constant-curvature completion in Eq.~(\ref{eq:lineelement}); this metric is not assumed to follow uniquely from the projective coordinate map. Two statements follow, and they are used repeatedly below. Equation~(\ref{eq:lineelement}) is an independent postulate: the FL projective map generates Eq.~(\ref{eq:pullback}), which is flat, and retaining only its conformal part is a separate modelling step. The map is also not a conformal transformation of Minkowski space, since Eq.~(\ref{eq:pullback}) is not proportional to $\eta_{\mu\nu}$. Whether the FL transformations act as isometries of Eq.~(\ref{eq:lineelement}) is not determined here, and the analysis does not depend on it.

This distinction also resolves the coordinate-artifact issue. A flat pullback can be removed globally by returning to $X^{\mu}$. The curvature of Eq.~(\ref{eq:lineelement}) cannot be removed by a coordinate transformation. Nevertheless, energies conjugate to the conformal time coordinate remain coordinate dependent, and their interpretation requires a local tetrad. Both statements are needed for a consistent analysis.

\subsection{Canonical phase space and mass shell}

The point-particle action on the background is
\be
 S_{\rm pp}=\int\dd\lambda\left[p_{\mu}\dot x^{\mu}
 -\frac{N}{2}\left(g^{\mu\nu}p_{\mu}p_{\nu}-m_{0}^{2}c^{2}\right)\right].
\label{eq:pp-action}
\ee
Its symplectic form and Poisson brackets are conventional,
\be
 \Omega=\dd p_{\mu}\wedge\dd x^{\mu},\qquad
 \{x^{\mu},p_{\nu}\}=\delta^{\mu}_{\nu}.
\label{eq:symplectic}
\ee
No momentum-space curvature, noncanonical bracket, or deformed position--momentum contraction is introduced. The mass-shell constraint is
\be
 g^{\mu\nu}p_{\mu}p_{\nu}=m_{0}^{2}c^{2}.
\label{eq:cov-massshell}
\ee
Since $g^{\mu\nu}=s^{2}\eta^{\mu\nu}$, setting $p_{0}=E/c$ gives Eq.~(\ref{eq:massshell-intro}). This equation is therefore an ordinary curved-spacetime relation expressed in conformal coordinates.

A local orthonormal observer uses the tetrad $e^{\hat a}{}_{\mu}=a\delta^{a}_{\mu}$. The locally measured energy and momentum are
\be
 E_{\rm loc}=c\,e_{\hat0}{}^{\mu}p_{\mu}=\frac{E}{a}=sE,
 \qquad p_{\hat i}=\frac{p_{i}}{a}.
\label{eq:local-energy}
\ee
Equation~(\ref{eq:cov-massshell}) then becomes the standard local relation
\be
 E_{\rm loc}^{2}-\vp_{\rm loc}^{2}c^{2}=m_{0}^{2}c^{4}.
\label{eq:local-massshell}
\ee
The apparent mass in conformal coordinates is therefore a tetrad effect rather than a varying invariant rest mass.

\section{Covariant Klein--Gordon field}
\label{sec:KG}

With the geometry and canonical phase space established, we now quantize the scalar sector directly from a covariant action, giving particular attention to the Hubble-type first-derivative term and to its elimination by a canonical field redefinition.

\subsection{Action and field equation}

For a complex scalar field, we consider
\be
 S_{\rm KG}=\int\dd^{4}x\sqrt{-g}\left[
 g^{\mu\nu}\partial_{\mu}\Phi^{*}\partial_{\nu}\Phi
 -\left(\frac{m_{0}^{2}c^{2}}{\hbar^{2}}-\xi\Ric\right)|\Phi|^{2}\right].
\label{eq:KG-action}
\ee
The sign of the curvature-coupling term is chosen consistently with Eq.~(\ref{eq:ricci}), so that $\xi=1/6$ is the conformal value in four dimensions. Variation gives
\be
 \left(\Box_{g}+\frac{m_{0}^{2}c^{2}}{\hbar^{2}}-\xi\Ric\right)\Phi=0,
\qquad
 \Box_{g}=\frac{1}{\sqrt{-g}}\partial_{\mu}
 \left(\sqrt{-g}g^{\mu\nu}\partial_{\nu}\right).
\label{eq:covKG}
\ee
In conformal time this becomes
\be
 \Phi''+2\Ham\Phi'-\nabla^{2}\Phi
 +a^{2}\left(\frac{m_{0}^{2}c^{2}}{\hbar^{2}}-\xi\Ric\right)\Phi=0.
\label{eq:KG-conformal-time}
\ee
Equivalently, in terms of $t$,
\be
 \ddot\Phi-\frac{2c}{Rs}\dot\Phi-c^{2}\nabla^{2}\Phi
 +\frac{1}{s^{2}}\left(\frac{m_{0}^{2}c^{4}}{\hbar^{2}}
 +\frac{12\xi c^{2}}{R^{2}}\right)\Phi=0.
\label{eq:KG-t}
\ee
The second term is the Hubble-type derivative term. Its sign reflects the decreasing scale factor in the chosen orientation of conformal time. Omitting it is inconsistent with the covariant action.

\subsection{Canonical field and curvature contribution}

The canonical conformal field is
\be
 \chi=a\Phi=\frac{\Phi}{s}.
\label{eq:chi-def}
\ee
After substituting $\Phi=\chi/a$ into the action and removing a total derivative, one obtains
\be
 S_{\rm KG}=\int\dd x^{0}\dd^{3}x
 \left[|\chi'|^{2}-|\nabla\chi|^{2}
 -\left(a^{2}\left(\frac{m_{0}^{2}c^{2}}{\hbar^{2}}-\xi\Ric\right)
 -\frac{a''}{a}\right)|\chi|^{2}\right].
\label{eq:chi-action}
\ee
The corresponding equation is
\be
 \chi''-\nabla^{2}\chi
 +\frac{1}{s^{2}}\left[
 \frac{m_{0}^{2}c^{2}}{\hbar^{2}}+\frac{12\xi-2}{R^{2}}
 \right]\chi=0.
\label{eq:chi-eq-x0}
\ee
In ordinary time,
\be
 \ddot\chi-c^{2}\nabla^{2}\chi
 +\frac{1}{s^{2}}\left[
 \frac{m_{0}^{2}c^{4}}{\hbar^{2}}+(12\xi-2)\frac{c^{2}}{R^{2}}
 \right]\chi=0.
\label{eq:chi-eq-t}
\ee
Thus the order-$R^{-1}$ first derivative is removable by a canonical field redefinition, but an order-$R^{-2}$ term remains. For a massless conformally coupled field, $m_{0}=0$ and $\xi=1/6$, the bracket vanishes, as required by conformal invariance. For minimal coupling, $\xi=0$, the shift is $-2c^{2}/(R^{2}s^{2})$. The two contributions to the coefficient $12\xi-2$ have different origins, a distinction used in Sec.~\ref{sec:operational}. The term $12\xi$ comes from the curvature coupling in the covariant action, Eq.~(\ref{eq:KG-action}), whereas the $-2$ is produced by the field redefinition $\chi=a\Phi$ through the term $-a''/a$ in Eq.~(\ref{eq:chi-action}). The second contribution is therefore a property of the chosen canonical variable and of the constant-$x^{0}$ slicing rather than an invariant feature of the field.

The momentum conjugate to $\chi$ is $\pi_{\chi}=\chi'^{*}$, and the equal-time commutator is
\be
 [\chi(x^{0},\vx),\pi_{\chi}(x^{0},\vy)]=i\hbar\delta^{3}(\vx-\vy).
\label{eq:KG-commutator}
\ee
The usual Fourier representation and the differential momentum operator follow directly from this canonical structure, providing the first-principles justification that a purely formal momentum-space substitution lacks.

\subsection{Exact modes and adiabatic regime}

For $\chi=\chi_{\vk}(x^{0})e^{i\vk\cdot\vx}$, Eq.~(\ref{eq:chi-eq-x0}) gives
\be
 \chi_{\vk}''+\omega_{k}^{2}(x^{0})\chi_{\vk}=0,
\qquad
 \omega_{k}^{2}=k^{2}+\frac{\kappa_{\xi}^{2}}{s^{2}},
\label{eq:modeeq}
\ee
where
\be
 \kappa_{\xi}^{2}=\frac{m_{0}^{2}c^{2}}{\hbar^{2}}+\frac{12\xi-2}{R^{2}}.
\label{eq:kappa}
\ee
For $k\neq0$, the exact solutions are
\be
 \chi_{\vk}(s)=\sqrt{s}\left[A_{\vk}H_{\nu}^{(1)}(kRs)
 +B_{\vk}H_{\nu}^{(2)}(kRs)\right],
\qquad
 \nu^{2}=\frac{1}{4}-\kappa_{\xi}^{2}R^{2}.
\label{eq:hankel}
\ee
The exact representation provides a useful check on the WKB expansion and makes explicit that the dynamics is standard field theory on a constant-curvature background.

In terms of $t$, the instantaneous frequency is
\be
 \Omega_{k}^{2}(t)=c^{2}k^{2}+\frac{c^{2}\kappa_{\xi}^{2}}{s^{2}(t)}.
\label{eq:Omega}
\ee
The adiabatic parameter is
\be
 \eta_{k}(t)=\left|\frac{\dot\Omega_{k}}{\Omega_{k}^{2}}\right|
 =\frac{c}{R}\frac{c^{2}\kappa_{\xi}^{2}/s^{3}}
 {\left(c^{2}k^{2}+c^{2}\kappa_{\xi}^{2}/s^{2}\right)^{3/2}}.
\label{eq:eta}
\ee
In the massive, low-momentum sector,
\be
 \eta_{k}\simeq\frac{1}{R\kappa_{\xi}}
 =\frac{\hbar}{m_{0}cR}\left[1+\mathcal O\left(\frac{\hbar^{2}}{m_{0}^{2}c^{2}R^{2}}\right)\right].
\label{eq:eta-mass}
\ee
For an electron and $R\sim c/H_{0}$, this ratio is of order $10^{-39}$. The conformal evolution is therefore extremely adiabatic on laboratory time scales.

Equation~(\ref{eq:eta}) is the WKB ratio of a second-order mode equation. It controls the accuracy of the positive- and negative-frequency basis, and hence the magnitude of the Bogoliubov mixing between those sectors~\cite{Birrell,ParkerToms,Parker1969,Fulling1973}, rather than a first-order transition probability. The same structure governs the trapped problem of Sec.~\ref{sec:KGosc}. The exact Hankel representation of Eq.~(\ref{eq:hankel}) makes an exact Bogoliubov computation available in the free sector, against which these adiabatic estimates can be checked.

\begin{figure}[t]
\centering
\includegraphics[width=0.76\linewidth]{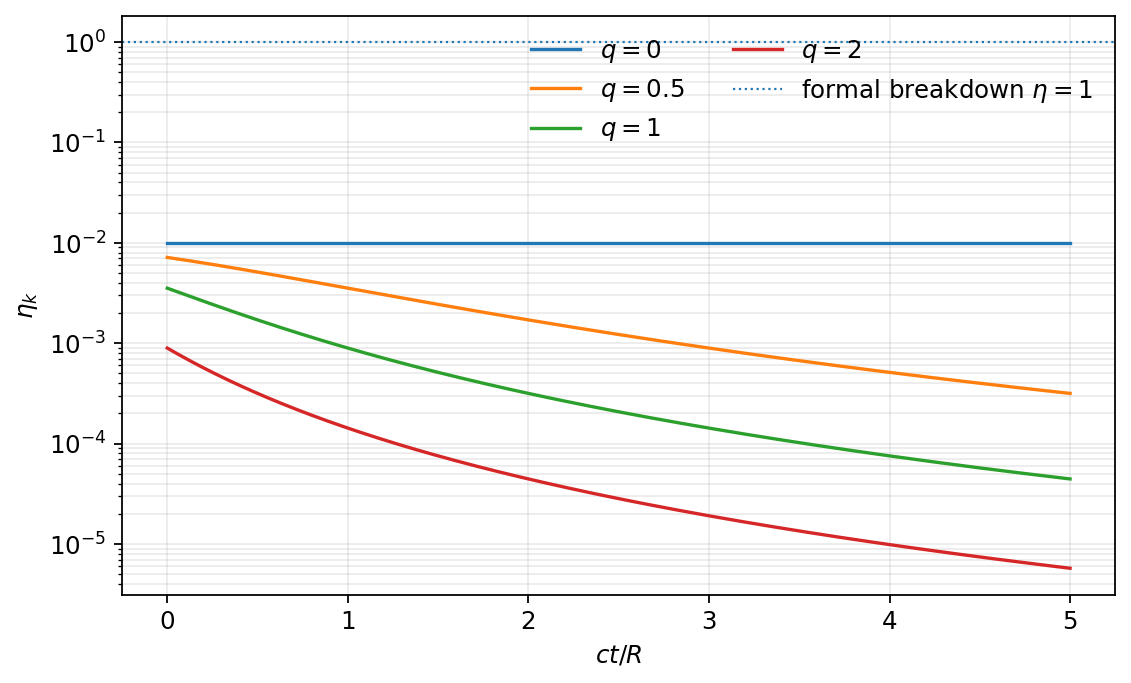}
\caption{Dimensionless adiabatic parameter for several values of $q=k/\kappa_{\xi}$, using the illustrative ratio $(R\kappa_{\xi})^{-1}=10^{-2}$. The zero-momentum curve is constant, while finite-momentum curves decrease as the kinetic term becomes dominant. Cosmological values of $R$ lie many orders of magnitude below the displayed scale.}
\label{fig:adiabatic}
\end{figure}

\subsection{Klein--Gordon inner product}

For two solutions of Eq.~(\ref{eq:covKG}), the conserved product is
\be
 (\Phi_{1},\Phi_{2})_{\rm KG}=i\int_{\Sigma}\dd\Sigma_{\mu}
 \left(\Phi_{1}^{*}\nabla^{\mu}\Phi_{2}
 -\Phi_{2}\nabla^{\mu}\Phi_{1}^{*}\right).
\label{eq:KG-product}
\ee
On a constant-$x^{0}$ surface this reduces, after Eq.~(\ref{eq:chi-def}), to the standard Wronskian product of the canonical fields. The field redefinition therefore changes the representation but not the physical solution space or its symplectic structure.

\section{Covariant Dirac field}
\label{sec:Dirac}

We now turn to the spinor sector, in which the geometry enters through the tetrad and the spin connection. The construction parallels the scalar case: a covariant action yields a first-order equation whose conformal factors are reorganized by a canonical spinor rescaling.

\subsection{Tetrad and spin connection}

The spinor action is
\be
 S_{D}=\int\dd^{4}x\sqrt{-g}\,
 \bar\psi\left[i\hbar c\gamma^{\mu}(x)\nabla_{\mu}-m_{0}c^{2}\right]\psi,
\label{eq:Dirac-action}
\ee
where $\gamma^{\mu}(x)=e_{a}{}^{\mu}\gamma^{a}$ and
\be
 \nabla_{\mu}=\partial_{\mu}+\Gamma_{\mu},\qquad
 \Gamma_{\mu}=\frac{1}{8}\omega_{\mu ab}[\gamma^{a},\gamma^{b}].
\label{eq:spin-cov}
\ee
A convenient diagonal tetrad is
\be
 e^{a}{}_{\mu}=a\delta^{a}_{\mu},\qquad
 e_{a}{}^{\mu}=a^{-1}\delta_{a}^{\mu}.
\label{eq:tetrad}
\ee
The spin connection gives the conformal Dirac equation
\be
 \left[i\hbar\gamma^{0}\left(\partial_{t}+\frac{3}{2}\frac{\dot a}{a}\right)
 +i\hbar c\gamma^{i}\partial_{i}-a m_{0}c^{2}\right]\psi=0.
\label{eq:dirac-physical}
\ee
The term $3\dot a/(2a)$ is the spinorial analogue of the scalar Hubble term.

\subsection{Canonical spinor and conformal mass}

Defining
\be
 \Psi=a^{3/2}\psi
\label{eq:spinor-rescale}
\ee
removes the spin connection from the first-order equation:
\be
 \left[i\hbar\gamma^{0}\partial_{t}+i\hbar c\gamma^{i}\partial_{i}
 -\mapp(t)c^{2}\right]\Psi=0,
\qquad
 \mapp(t)=a(t)m_{0}.
\label{eq:dirac-rescaled}
\ee
The Hamiltonian is
\be
 H_{D}(t)=c\boldsymbol{\alpha}\cdot\vp+\beta\mapp(t)c^{2}.
\label{eq:dirac-H}
\ee
The conserved curved-space inner product becomes
\be
 (\psi_{1},\psi_{2})_{D}=\int\dd^{3}x\,a^{3}\psi_{1}^{\dagger}\psi_{2}
 =\int\dd^{3}x\,\Psi_{1}^{\dagger}\Psi_{2}.
\label{eq:dirac-inner}
\ee
Hence $H_{D}(t)$ is Hermitian in the canonical Hilbert space of $\Psi$. The ordinary replacement $\vp=-i\hbar\nabla$ is here a derived consequence of the action and tetrad structure rather than a postulate imposed through a deformed mass shell.

\subsection{Squared operator and geometric correction}

Multiplying Eq.~(\ref{eq:dirac-rescaled}) by the conjugate first-order operator gives
\be
 \left[\partial_{t}^{2}-c^{2}\nabla^{2}
 +\frac{\mapp^{2}c^{4}}{\hbar^{2}}
 +i\frac{c^{2}}{\hbar}\dot\mapp\gamma^{0}\right]\Psi=0.
\label{eq:dirac-square}
\ee
The correction is
\be
 \Delta_{D}=i\frac{c^{2}}{\hbar}\dot\mapp\gamma^{0},
\qquad
 \dot\mapp=-\frac{c}{Rs}\mapp.
\label{eq:DeltaD}
\ee
Its magnitude relative to the mass term is exactly
\be
 \frac{|\Delta_{D}|}{\mapp^{2}c^{4}/\hbar^{2}}
 =\frac{\hbar}{m_{0}cR}.
\label{eq:dirac-ratio}
\ee
The derivative term is therefore the conformal image of the curved-spacetime spin connection. In the massless limit the rescaled Dirac equation is exactly flat, reflecting the conformal invariance of the four-dimensional massless Dirac field.

\section{Klein--Gordon oscillator}
\label{sec:KGosc}

\subsection{Status of the oscillator coupling}
\label{sec:osc-status}

The oscillator interactions used in this section and in Sec.~\ref{sec:Diracosc} are introduced differently from the free field equations obtained above. This distinction must be stated before computing the spectra.

The free scalar and spinor equations follow from the covariant actions of Eqs.~(\ref{eq:KG-action}) and~(\ref{eq:Dirac-action}). The oscillator couplings are introduced instead at the level of the canonical equations, after the conformal field rescalings of Eqs.~(\ref{eq:chi-def}) and~(\ref{eq:spinor-rescale}), through the nonminimal substitutions of Eqs.~(\ref{eq:KG-partners}) and~(\ref{eq:DO-sub}). The background geometry does not determine three ingredients of this construction: the conformal coordinate $x$ in which the potential is taken to be quadratic, the coefficient $\mapp(t)$ multiplying it, and the transformation rule adopted for the trapping frequency $\omega_{c}$. The construction does not include a covariant scalar, vector, electromagnetic or tetrad coupling that generates this trap.

The Klein--Gordon and Dirac oscillators of Secs.~\ref{sec:KGosc} and~\ref{sec:Diracosc} are therefore \emph{coordinate-defined simplified models}. Their fixed-time eigenvalues are eigenvalues of chart-dependent operators: parameters of the model rather than energies measured by a device, and not components of a particle four-momentum. This qualification applies to all spectra discussed below.

A covariantly defined trap requires additional ingredients. One would build Fermi normal coordinates along the world line of a chosen observer~\cite{ManasseMisner}; specify a trapping field---a static scalar profile or an electromagnetic configuration, for instance---whose parameters are fixed in the observer's local frame by a physical clock; couple it covariantly to the quantum field; and obtain the response of a detector with a specified trajectory, switching function and interaction~\cite{LoukoSatz2008}. Such a construction also generates tidal terms of relative order $\ell_{\rm lab}^{2}/R^{2}$, which the models below do not contain. It lies outside the scope of the present analysis, and the oscillators treated here are defined directly in the conformal chart.

\subsection{Partner operators and ordering structure}

The one-dimensional Klein--Gordon oscillator is introduced in the canonical scalar equation, whose kinetic term and inner product have standard form. The original construction admits two adjoint factor orderings that form a supersymmetric partner pair~\cite{Bruce,Junker2021}. With
\be
 \Lambda(t)=\mapp(t)\omega_{c}(t),\qquad p_{x}=-i\hbar\partial_{x},
\label{eq:Lambda}
\ee
we define
\bea
 \mathcal O_{0}&=&(p_{x}+i\Lambda x)(p_{x}-i\Lambda x)
 =p_{x}^{2}+\Lambda^{2}x^{2}-\hbar\Lambda,\nonumber\\
 \mathcal O_{1}&=&(p_{x}-i\Lambda x)(p_{x}+i\Lambda x)
 =p_{x}^{2}+\Lambda^{2}x^{2}+\hbar\Lambda.
\label{eq:KG-partners}
\eea
The corresponding canonical equations are
\be
 \left[\hbar^{2}\partial_{t}^{2}+c^{2}\mathcal O_{\rho}
 +\mapp^{2}c^{4}+\Delta_{\xi}(t)\right]\chi_{\rho}=0,
 \qquad \rho=0,1,
\label{eq:KGosc}
\ee
where
\be
 \Delta_{\xi}(t)=(12\xi-2)\frac{\hbar^{2}c^{2}}{R^{2}s^{2}(t)}.
\label{eq:Uxi}
\ee
The sector $\rho=0$ has an unshifted oscillator ground level, whereas $\rho=1$ is shifted upward by one quantum. Retaining both sectors avoids identifying a particular factor ordering with the complete Klein--Gordon oscillator spectrum.

\subsection{Instantaneous spectra and eigenfunctions}

At fixed time, the eigenvalues of $\mathcal O_{\rho}$ are
\be
 \lambda_{n,\rho}=2\hbar\Lambda(t)(n+\rho),
 \qquad n=0,1,2,\ldots .
\label{eq:KG-partner-eigenvalues}
\ee
The leading adiabatic coordinate-time spectra are therefore
\be
 E_{n,\rho,c}^{2}(t)=\mapp^{2}(t)c^{4}
 +2\mapp(t)c^{2}\hbar\omega_{c}(t)(n+\rho)
 +(12\xi-2)\frac{\hbar^{2}c^{2}}{R^{2}s^{2}(t)},
\label{eq:KG-spectrum}
\ee
with $\rho=0,1$. The normalized instantaneous eigenfunctions are Hermite functions with the replacement $m_{0}\omega\rightarrow\mapp\omega_{c}$. Both partner sectors have the same coordinate oscillator length,
\be
 \ell_{c}(t)=\sqrt{\frac{\hbar}{\mapp(t)\omega_{c}(t)}}.
\label{eq:KG-length}
\ee
The order-$R^{-2}$ term shifts the canonical instantaneous eigenvalues but does not modify the spatial eigenfunctions at this order.

\begin{figure}[t]
\centering
\includegraphics[width=0.76\linewidth]{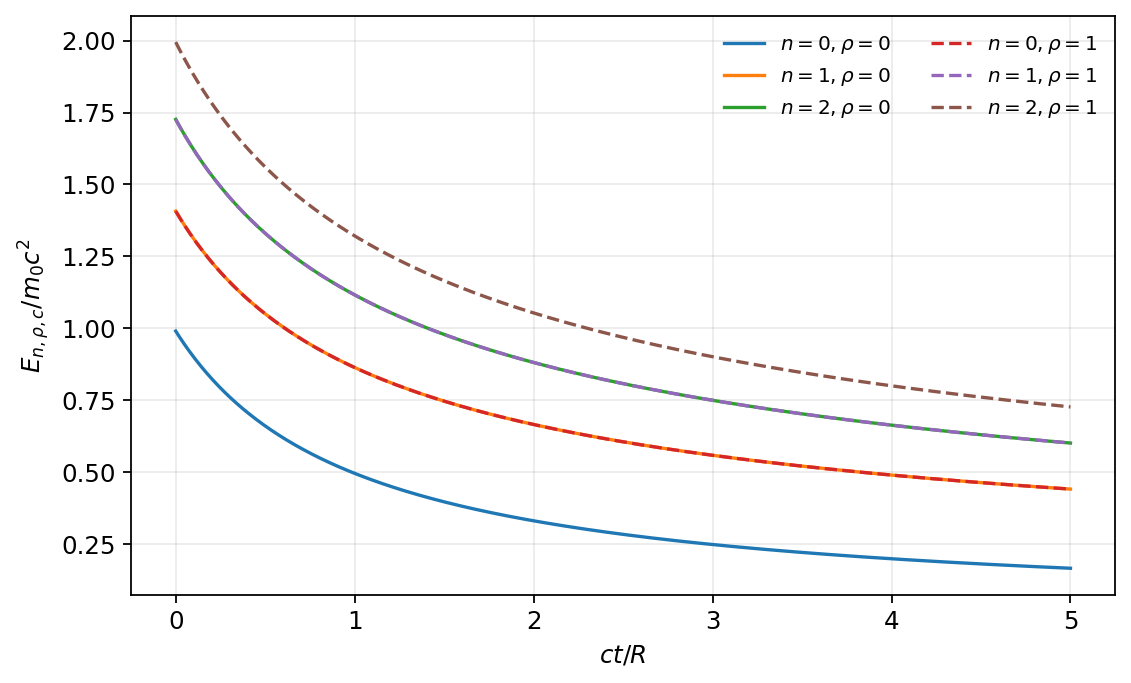}
\caption{Positive coordinate-time Klein--Gordon oscillator energies for the two partner sectors $\rho=0$ (solid) and $\rho=1$ (dashed), at fixed $\hbar\omega_{c}/(m_{0}c^{2})=0.5$. The curvature coefficient is set to $(12\xi-2)[\hbar/(m_{0}cR)]^{2}=-0.02$ for visibility.}
\label{fig:KG-spectrum}
\end{figure}

For fixed coordinate frequency, $\omega_{c}=\omega_{0}$, Eq.~(\ref{eq:KG-spectrum}) displays the conformal-coordinate drift of both partner ladders. Figure~\ref{fig:KG-spectrum} uses an enlarged curvature parameter solely to make the small shift visible. As established in Sec.~\ref{sec:operational}, this drift is a property of the chart and of the frequency prescription.

\subsection{Second-order adiabatic analysis}
\label{sec:KG-adiabatic}

The canonical Klein--Gordon equation~(\ref{eq:KGosc}) is second order in time, and its solution space carries the Klein--Gordon product of Eq.~(\ref{eq:KG-product}) rather than an $L^{2}$ norm. Its adiabatic analysis is therefore organized as a second-order problem with explicit positive- and negative-frequency sectors; the first-order adiabatic transition formula of a Schr\"odinger equation $i\hbar\partial_{t}\psi=H(t)\psi$ does not apply to it.

The instantaneous Hermite basis $\phi_{n,\rho}(x;t)$ depends on time through $\Lambda(t)=\mapp(t)\omega_{c}(t)$. Differentiation couples only states of equal parity, and up to phase conventions
\bea
 \left|\langle n+2|\partial_{t}n\rangle\right|
 &=&\frac{|\dot\Lambda|}{4\Lambda}\sqrt{(n+1)(n+2)},\nonumber\\
 \left|\langle n-2|\partial_{t}n\rangle\right|
 &=&\frac{|\dot\Lambda|}{4\Lambda}\sqrt{n(n-1)}.
\label{eq:KG-couplings}
\eea
These are basis-derivative matrix elements; their role in the mixing of the second-order system is derived below.

Expanding the canonical field in the instantaneous basis,
\be
 \chi(t,x)=\sum_{n}q_{n}(t)\,\phi_{n}(x;t),
\label{eq:KG-expansion}
\ee
and substituting into Eq.~(\ref{eq:KGosc}) gives a coupled system of \emph{second-order} amplitude equations,
\be
 \ddot q_{m}+\frac{E_{m}^{2}(t)}{\hbar^{2}}\,q_{m}
 =-\sum_{n}\left[\,2\langle m|\partial_{t}n\rangle\,\dot q_{n}
 +\langle m|\partial_{t}^{2}n\rangle\,q_{n}\right],
\label{eq:KG-amplitude-system}
\ee
in which the partner index $\rho$ is suppressed. Dropping the right-hand side leaves the equation with fixed parameters $\ddot q_{m}+\hbar^{-2}E_{m}^{2}(t)q_{m}=0$, to which the second-order WKB positive- and negative-frequency basis applies.

Three independent small parameters control the approximation, each defined using the relevant instantaneous gap. The first is the WKB ratio of a given level,
\be
 \eta_{m}(t)=\hbar\left|\frac{\dot E_{m}(t)}{E_{m}^{2}(t)}\right|,
\label{eq:eta-level}
\ee
which governs the mixing between the positive- and negative-frequency solutions of the \emph{same} level, whose separation is the particle--antiparticle gap $2E_{m}(t)$. The second is the ratio of the basis-derivative couplings to the gaps between distinct levels,
\be
 \zeta_{mn}(t)=\frac{\hbar\left|\langle m|\partial_{t}n\rangle\right|}
 {\left|E_{m}(t)-E_{n}(t)\right|},
\label{eq:zeta}
\ee
and the third is the corresponding ratio for the pair gap,
\be
 \widetilde\zeta_{mn}(t)=\frac{\hbar\left|\langle m|\partial_{t}n\rangle\right|}
 {E_{m}(t)+E_{n}(t)},
\label{eq:zeta-tilde}
\ee
which measures the pair-creation channel. Adiabaticity requires all three to be small. The resulting mixing is expressed either through Bogoliubov coefficients or, equivalently, through the two-component Feshbach--Villars form of the Klein--Gordon equation, in which first-order methods apply because the positive- and negative-frequency components are carried explicitly~\cite{FeshbachVillars,Birrell,ParkerToms,Parker1969,Fulling1973}.

Evaluating the three parameters for the present model separates two distinct scales. For a fixed coordinate frequency, $\dot\Lambda/\Lambda=\dot\mapp/\mapp=-c/(Rs)$. In the regime $\hbar\omega_{c}\ll\mapp c^{2}$ the ladders of Eq.~(\ref{eq:KG-spectrum}) give $E_{m}-E_{n}\simeq\hbar\omega_{c}(m-n)$ and $E_{m}+E_{n}\simeq2\mapp c^{2}$, so that, up to factors involving the level index,
\be
 \zeta\sim\frac{c}{R\,s(t)\,\omega_{0}},
\qquad
 \eta_{m}\simeq\widetilde\zeta\sim\frac{\hbar}{m_{0}cR}.
\label{eq:zeta-estimate}
\ee
The level-mixing channel is thus controlled by the ratio of the FL time scale to the oscillator period, whereas the positive/negative-frequency mixing of a given level and the pair-creation channel are controlled by the ratio of the Compton wavelength to the curvature radius. For an electron with $R\simeq c/H_{0}$ the two differ by about fifteen orders of magnitude, so that $c/(Rs\omega_{0})$ measures the separation of time scales alone and is not by itself the adiabatic parameter of the problem.
Under the fixed local-frequency prescription of Sec.~\ref{sec:operational} one has $\Lambda=m_{0}a^{2}\omega_{0}$, so that $|\dot\Lambda/\Lambda|$ is twice the value above and all three parameters are rescaled by factors of order unity. For microscopic frequencies and a cosmological curvature radius each of them is negligible, and the spectra computed at fixed parameters are internally consistent within the chart-defined model.

\section{Dirac oscillator}
\label{sec:Diracosc}

\subsection{Hamiltonian and component equations}

The Dirac oscillator is defined in the canonical spinor equation through the nonminimal substitution~\cite{Moshinsky,Martinez1995}
\be
 p_{x}\rightarrow p_{x}-i\mapp(t)\omega_{c}(t)x\beta.
\label{eq:DO-sub}
\ee
In $1+1$ dimensions we take $\gamma^{0}=\sigma_{3}$, $\gamma^{1}=i\sigma_{2}$, and $\alpha=\sigma_{1}$. The Hamiltonian is
\be
 H_{\rm DO}(t)=c\alpha\left[p_{x}-i\mapp\omega_{c}x\beta\right]
 +\beta\mapp c^{2}.
\label{eq:DO-H}
\ee
In natural units, for $\Psi=(\psi_{1},\psi_{2})^{T}$,
\be
 H_{\rm DO}=
 \begin{pmatrix}
 \mapp & -i\partial_{x}+i\mapp\omega_{c}x\\
 -i\partial_{x}-i\mapp\omega_{c}x & -\mapp
 \end{pmatrix}.
\label{eq:DO-matrix}
\ee
At fixed time, define $A=-i\partial_{x}+i\mapp\omega_{c}x$ and $B=-i\partial_{x}-i\mapp\omega_{c}x$. Their products are
\be
 AB=-\partial_{x}^{2}+\mapp^{2}\omega_{c}^{2}x^{2}-\mapp\omega_{c},
 \qquad
 BA=-\partial_{x}^{2}+\mapp^{2}\omega_{c}^{2}x^{2}+\mapp\omega_{c}.
\label{eq:ABBA}
\ee

\subsection{Energy branches and isolated state}

The upper and lower component equations yield
\bea
 E_{n,+}^{2}&=&\mapp^{2}c^{4}+2\mapp c^{2}\hbar\omega_{c}n,\nonumber\\
 E_{n,-}^{2}&=&\mapp^{2}c^{4}+2\mapp c^{2}\hbar\omega_{c}(n+1).
\label{eq:DO-components}
\eea
For $N\geq1$, a physical spinor level pairs the upper component with index $N$ and the lower component with index $N-1$. The positive- and negative-energy branches are
\be
 E_{N,c}^{(\pm)}(t)=\pm\sqrt{\mapp^{2}(t)c^{4}
 +2\mapp(t)c^{2}\hbar\omega_{c}(t)N},
 \qquad N=1,2,\ldots .
\label{eq:DO-spectrum}
\ee
For the sign convention in Eq.~(\ref{eq:DO-sub}), the unpaired level is
\be
 E_{0,c}(t)=+\mapp(t)c^{2}.
\label{eq:DO-isolated}
\ee
Reversing the sign of the oscillator coupling exchanges this isolated positive-energy state with an isolated negative-energy state. The squared spectrum alone therefore does not display the spectral asymmetry.

\begin{figure}[t]
\centering
\includegraphics[width=0.76\linewidth]{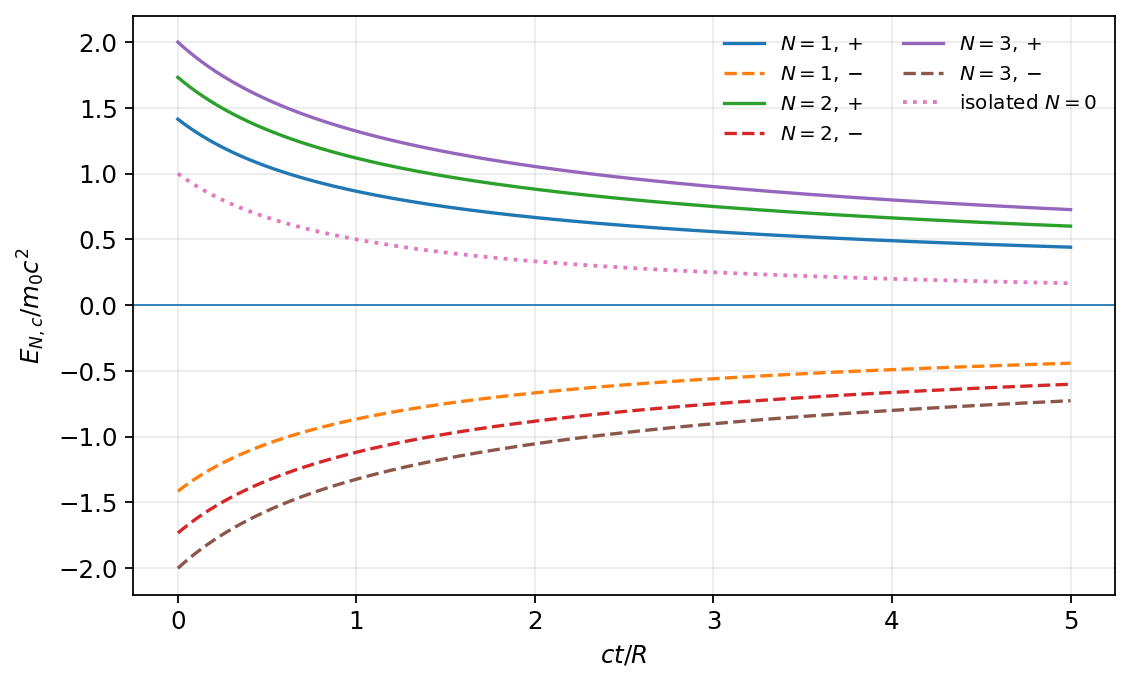}
\caption{Positive- and negative-energy Dirac-oscillator branches for fixed coordinate frequency. The isolated $N=0$ state occurs only on the positive branch for the coupling convention used in Eq.~(\ref{eq:DO-sub}).}
\label{fig:DO-spectrum}
\end{figure}

\subsection{Time-dependent factorization and spinor mixing}

For an explicitly time-dependent Hamiltonian, direct operator multiplication gives
\be
 (i\hbar\partial_{t}+H_{\rm DO})(i\hbar\partial_{t}-H_{\rm DO})\Psi
 =\left[-\hbar^{2}\partial_{t}^{2}-i\hbar\dot H_{\rm DO}
 -H_{\rm DO}^{2}\right]\Psi.
\label{eq:DO-square-time}
\ee
Thus every solution of $(i\hbar\partial_{t}-H_{\rm DO})\Psi=0$ obeys
\be
 \left[\hbar^{2}\partial_{t}^{2}+i\hbar\dot H_{\rm DO}
 +H_{\rm DO}^{2}\right]\Psi=0.
\label{eq:DO-second-order}
\ee
The derivative $\dot H_{\rm DO}$ contains the mass contribution $\beta c^{2}\dot\mapp$ and the derivative of $\mapp\omega_{c}$. These terms generate nonadiabatic mixing between instantaneous oscillator levels and between spinor components. The Dirac sector differs structurally from the Klein--Gordon sector in this respect: Eq.~(\ref{eq:dirac-rescaled}) is first order in time and $H_{\rm DO}(t)$ is Hermitian in the canonical inner product of Eq.~(\ref{eq:dirac-inner}), so the standard adiabatic theorem applies once the matrix elements are compared with the relevant instantaneous gaps. Two kinds of gap occur: the spacing within a branch $E_{N+1,c}^{(+)}-E_{N,c}^{(+)}$, and the particle--antiparticle gap $E_{N,c}^{(+)}-E_{N,c}^{(-)}=2E_{N,c}^{(+)}$, which governs the pair channel and the Zitterbewegung-type component mixing. The two entries of the estimate below correspond to these two channels and are the Dirac counterparts of Eq.~(\ref{eq:zeta-estimate}). A useful fixed-coordinate-frequency estimate is
\be
 \epsilon_{D}(t)\sim\max\left(
 \frac{c}{Rs(t)\omega_{c}(t)},
 \frac{\hbar}{m_{0}cR}\right).
\label{eq:epsilonD}
\ee
For the fixed local-frequency prescription, the logarithmic derivative of $\mapp\omega_{c}=m_{0}a^{2}\omega_{0}$ introduces the expected factor of two in the basis-variation term. Both contributions are negligible for microscopic systems when $R$ is cosmological. They describe dynamical evolution of the model, whereas the leading coordinate-energy drift is removed by the change of prescription and rescaling treated in Sec.~\ref{sec:operational}.

\section{Coordinate spectra and operational observables}
\label{sec:operational}

The relation between coordinate-dependent spectral behaviour and operationally defined local physics is examined by comparing two prescriptions for the trapping frequency. The comparison is algebraic, is carried out at leading adiabatic order, and applies to the coordinate-defined models of Secs.~\ref{sec:KGosc} and~\ref{sec:Diracosc}; the quantities it produces are eigenvalues of chart-dependent operators rather than detector readings. It identifies the part of the spectral drift that can be removed by changing the frequency prescription and rescaling the eigenvalue.

\subsection{Coordinate and local frequency protocols}

An energy $E_{c}$ conjugate to conformal coordinate time is related to the tetrad-projected local energy by Eq.~(\ref{eq:local-energy}), $E_{\rm loc}=E_{c}/a$. That relation concerns the components of a point-particle four-momentum covector at an event; its extension to eigenvalues of a time-dependent field operator is examined at the end of this subsection. The frequency parameter of an externally maintained oscillator is taken to transform in the same manner, which forms part of the definition of the model. Two prescriptions must be distinguished.

For a fixed coordinate frequency,
\be
 \omega_{c}(t)=\omega_{0},
\label{eq:coord-protocol}
\ee
the corresponding local frequency is $\omega_{\rm loc}=\omega_{c}/a=s\omega_{0}$. The device is therefore not stationary in proper time, and its conformal-coordinate spectra drift.

For a fixed local frequency,
\be
 \omega_{\rm loc}=\omega_{0},\qquad
 \omega_{c}(t)=a(t)\omega_{0}.
\label{eq:local-protocol}
\ee
Substitution into the Klein--Gordon partner spectra gives
\be
 E_{n,\rho,c}^{2}=a^{2}\left[m_{0}^{2}c^{4}
 +2m_{0}c^{2}\hbar\omega_{0}(n+\rho)
 +(12\xi-2)\frac{\hbar^{2}c^{2}}{R^{2}}\right].
\label{eq:KG-local-protocol-coordinate}
\ee
Dividing by the scale factor, as for the components of a point-particle covector, defines the rescaled quantity
\be
 \widetilde E_{n,\rho}^{\,2}(t)\equiv\frac{E_{n,\rho,c}^{2}(t)}{a^{2}(t)}
 =m_{0}^{2}c^{4}
 +2m_{0}c^{2}\hbar\omega_{0}(n+\rho)
 +(12\xi-2)\frac{\hbar^{2}c^{2}}{R^{2}},
\label{eq:KG-local-spectrum}
\ee
in which the time-dependent conformal factor has cancelled.

Equation~(\ref{eq:KG-local-spectrum}) is not a locally measured spectrum, for three independent reasons. The tetrad relation $E_{\rm loc}=E_{c}/a$ of Eq.~(\ref{eq:local-energy}) concerns the components of a point-particle four-momentum covector at an event, whereas Eq.~(\ref{eq:KG-local-spectrum}) applies it to an instantaneous eigenvalue of a time-dependent, conformally rescaled \emph{field} operator, and this extension is not justified by the tetrad relation alone.

First, the background is nonstationary: $\partial_{t}$ is not a Killing vector of Eq.~(\ref{eq:lineelement}), so no conserved Killing energy exists of which the eigenvalue could be a tetrad component. Second, an instantaneous eigenfrequency of the canonical field $\chi$ is not in general a component of a particle four-momentum; such an identification requires a particle interpretation, which a time-dependent background fixes only through a choice of vacuum and which is ambiguous at the order of the terms under discussion. Third, the term proportional to $-2/R^{2}$ in Eq.~(\ref{eq:KG-local-spectrum}) originates not in the geometry but, as shown after Eq.~(\ref{eq:chi-eq-t}), in the field redefinition $\chi=a\Phi$ through $-a''/a$; it therefore depends on the field representation and the choice of time slices, and division of a coordinate eigenvalue by $a^{2}$ does not convert it into a local level shift. Time-dependent field redefinitions, positive/negative-frequency mixing and the detector coupling all contribute at this order.

Equation~(\ref{eq:KG-local-spectrum}) therefore supports one statement, which is the principal result of this section: the entire leading time dependence of Eq.~(\ref{eq:KG-spectrum}) under the fixed-coordinate-frequency prescription is removed by a change of prescription together with a rescaling. A drift removable in this way follows from the chart and from the prescription rather than from the dynamical content of the model, and this conclusion does not require the rescaled quantity to be an observable.

For the Dirac oscillator the same prescription gives, for $N\geq1$, the rescaled quantity
\be
 \widetilde E_{N}^{(\pm)}=\pm\sqrt{m_{0}^{2}c^{4}
 +2m_{0}c^{2}\hbar\omega_{0}N},
\label{eq:DO-local-spectrum}
\ee
with the isolated level $\widetilde E_{0}=+m_{0}c^{2}$ for the coupling sign used here. This coincides formally with the flat-space Dirac-oscillator spectrum, and the qualification stated above applies unchanged: Eq.~(\ref{eq:DO-local-spectrum}) is a rescaled eigenvalue of the chart-defined Hamiltonian, and it shows that the leading drift can be removed in the same way. Dynamical effects enter instead through the nonadiabatic spinor mixing of Sec.~\ref{sec:Diracosc} and through tidal terms in a finite Fermi-normal laboratory, neither of which is contained in Eq.~(\ref{eq:DO-local-spectrum}).

\begin{figure}[t]
\centering
\includegraphics[width=0.78\linewidth]{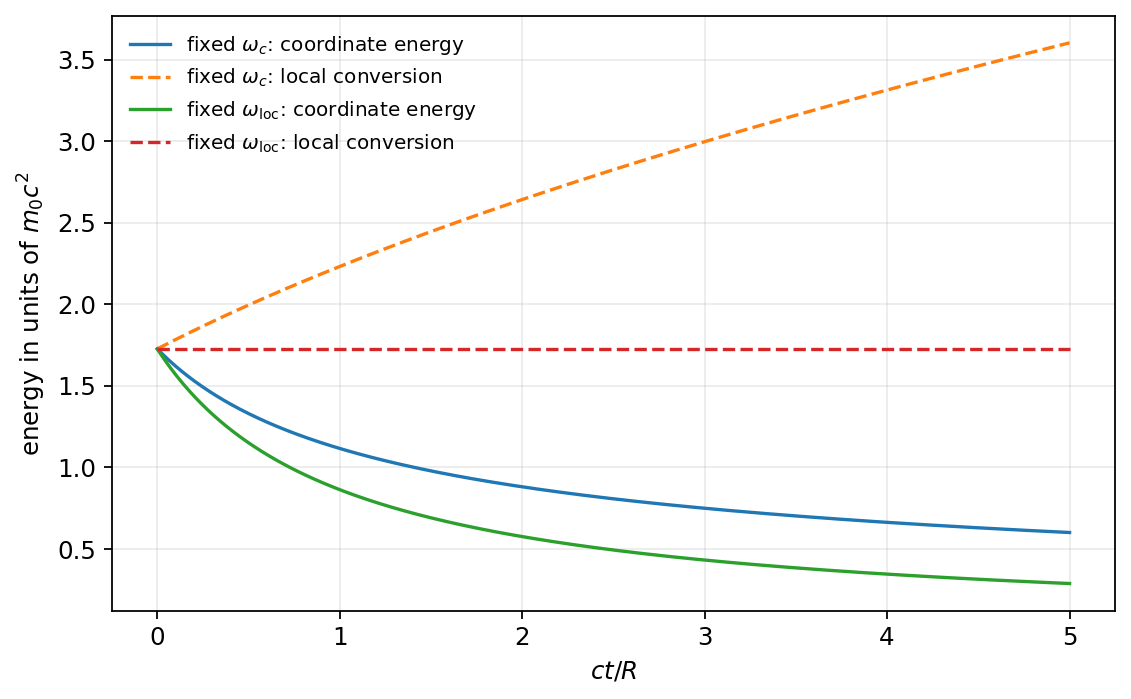}
\caption{Coordinate and tetrad-converted adiabatic energies for the $n=1$, $\rho=1$ Klein--Gordon partner level. A fixed coordinate frequency defines a time-dependent proper trap. For fixed local frequency, the coordinate energy scales with $a(t)$ and the tetrad-converted energy is constant at the displayed order.} The rescaled curve is the quantity $\widetilde E$ of Eq.~(\ref{eq:KG-local-spectrum}), a rescaled coordinate eigenvalue rather than a measured local energy.
\label{fig:protocol}
\end{figure}

\subsection{Localization length}

The physical oscillator length is $\ell_{\rm phys}=a\ell_{c}$. For fixed coordinate frequency,
\be
 \ell_{\rm phys}\propto a^{1/2},
\label{eq:length-coordinate}
\ee
so the proper size of the trap changes. For fixed local frequency, $\mapp\omega_{c}=a^{2}m_{0}\omega_{0}$ and
\be
 \ell_{c}=\frac{1}{a}\sqrt{\frac{\hbar}{m_{0}\omega_{0}}},\qquad
 \ell_{\rm phys}=\sqrt{\frac{\hbar}{m_{0}\omega_{0}}}.
\label{eq:length-local}
\ee
The coordinate broadening of the wavefunction is therefore compensated by the conversion from coordinate to proper length.

\subsection{Effects that can survive the rescaling}
\label{sec:residual}

Removing the leading drift does not account for all physical effects of the geometry. The effects that can survive are of three kinds, and each requires structure beyond the models solved above. Tidal corrections of relative order $\ell_{\rm lab}^{2}/R^{2}$ require a finite laboratory constructed in Fermi normal coordinates~\cite{ManasseMisner}, and are absent from a model whose trap is defined in the conformal chart. Nonuniversal curvature couplings require a specification of how each microscopic sector couples to $\Ric$. Nonadiabatic effects---level mixing and particle production---are present in the model and are controlled by the parameters of Eqs.~(\ref{eq:eta-level})--(\ref{eq:zeta-tilde}), although a fixed-time spectrum does not exhibit them.

An observable statement must in addition be dimensionless. A ratio such as
\be
 \mathcal Q_{AB}=\frac{\nu_{A}^{\rm loc}}{\nu_{B}^{\rm loc}},
\label{eq:ratio}
\ee
with both clocks and their microscopic couplings specified, is the appropriate quantity: universal conformal factors cancel from it, while nonuniversal curvature couplings or tidal responses need not. Computing such a ratio requires the trapping field, clock and detector described in Sec.~\ref{sec:osc-status}.

\section{Relation to nonlinear-relativity frameworks}
\label{sec:comparison}

The constant-curvature FL model is structurally distinct from DSR, gravity's rainbow, and relative locality. Its defining ingredients are summarized in Table~\ref{tab:comparison}.

\begin{table}[t]
\caption{Structural comparison of the constant-curvature FL model with nonlinear-relativity alternatives.}
\label{tab:comparison}
\resizebox{\textwidth}{!}{%
\begin{tabular}{llll}
\toprule
Framework & Momentum sector & Spacetime sector & Quantization structure\\
\midrule
Constant-curvature FL model & Canonical cotangent fibers & One energy-independent de Sitter metric & Covariant KG and Dirac actions\\
Gravity's rainbow & Nonlinear energy-momentum map & Energy-dependent metric & Mode-dependent prescription\\
Nonlinear position relativity & Nonlinear Lorentz action & Energy dependence or nonlinear pairing & Deformed plane-wave structure\\
Relative locality & Curved momentum space & Phase-space-dependent localization & Momentum-geometry-based\\
\bottomrule
\end{tabular}}
\end{table}

The two alternatives in Ref.~\cite{MSPosition} explain why a formal position--momentum exchange is insufficient. If the linear phase $p_{a}x^{a}$ is retained, the spacetime transformation becomes energy dependent. If position space is assigned an independent nonlinear transformation, the invariant pairing must be modified. The constant-curvature FL theory adopts neither option. It retains the standard pairing and introduces a single spacetime metric that is independent of the particle energy.

The classification in Table~\ref{tab:comparison} is not exhaustive. Nonlinear kinematic variables do not in general require a deformed symplectic structure, a noncanonical position--momentum pairing or a modified dispersion relation: minimum-length constructions have been given in which position, time, energy and momentum operators are all modified, while the canonical commutator and, in the relevant formulation, the standard dispersion relation are retained~\cite{Bishop2021,Bishop2022}. The table entries therefore apply to the specific frameworks listed. The present construction avoids phase-space deformation by a different route, placing the nonlinearity in a fixed, energy-independent spacetime metric.

The mass shell in Eq.~(\ref{eq:massshell-intro}) should therefore not be described as a deformed Casimir of a new relativity group. It is the coordinate representation of Eq.~(\ref{eq:cov-massshell}). Likewise, $\mapp(t)$ is not a varying invariant mass. It is the mass coefficient of canonically rescaled fields in conformal time. This distinction determines the appropriate actions, inner products, and observable quantities.

\section{Physical scales, limitations, and further developments}
\label{sec:limits}

The geometric corrections are suppressed by powers of $R^{-1}$. For $R\simeq c/H_{0}$, the scalar curvature scale is $|\Ric|^{1/2}\sim H_{0}/c$. The relative scalar mass correction is
\be
 \left|12\xi-2\right|\left(\frac{\hbar}{m_{0}cR}\right)^{2}.
\label{eq:scalar-relative}
\ee
For an electron this is of order $10^{-77}$, while the first-order Dirac adiabatic ratio in Eq.~(\ref{eq:dirac-ratio}) is of order $10^{-39}$. Oscillator nonadiabaticity is controlled by $H_{0}/\omega_{0}$ and is of order $10^{-24}$ for megahertz devices. Laboratory observation is therefore not expected if $R$ is cosmological.

The analysis is subject to the following limitations, of which the first two are structural.

First, Eq.~(\ref{eq:lineelement}) is a constant-curvature completion of the FL conformal factor and not a consequence of the FL transformations, whose projective map pulls Minkowski space back to the flat metric of Eq.~(\ref{eq:pullback}). The field-theoretic content of Secs.~\ref{sec:geometry}--\ref{sec:Dirac} is de Sitter field theory written in the FL parametrization; the FL-specific content is the chart, the coefficient $\mapp(t)$ it generates, and the operational question treated in Sec.~\ref{sec:operational}.

Second, the oscillator interaction is defined in the conformal chart and is not generated by a covariant trapping field, as set out in Sec.~\ref{sec:osc-status}. Both oscillator sectors are coordinate-defined simplified models, their fixed-time eigenvalues are chart-dependent parameters, and no quantity computed from them is identified with a measured local energy.

Third, the Klein--Gordon treatment is adiabatic. The exact Hankel modes of Eq.~(\ref{eq:hankel}) permit an exact Bogoliubov analysis of the free sector, while the trapped problem is treated through the second-order estimates of Sec.~\ref{sec:KG-adiabatic}. Fourth, the oscillator sectors are restricted to one spatial dimension, so spin--orbit structure and degeneracy splitting are not addressed. Fifth, quantum backreaction on the background is neglected.

Two extensions follow directly from the first two limitations. A covariant laboratory model would be constructed in Fermi normal coordinates along a chosen world line~\cite{ManasseMisner}, with the oscillator potential generated by a scalar or electromagnetic field whose local frequency is fixed by a dynamical clock, and with the measured quantity defined through a detector model of specified trajectory, switching and coupling~\cite{LoukoSatz2008}. Such a construction would separate $R^{-2}$ tidal corrections, canonical field-redefinition terms and genuine curvature-coupling effects. A second extension is the exact Bogoliubov analysis of the modes in Eq.~(\ref{eq:hankel}), together with a comparison of adiabatic vacua, which would fix the particle content of the free sector unambiguously. Both are well defined within this geometry and neither requires a deformed phase-space algebra.

\section{Conclusion}
\label{sec:conclusion}

Klein--Gordon and Dirac oscillators have been formulated on the constant-curvature Fock--Lorentz background $g_{\mu\nu}=s^{-2}(t)\eta_{\mu\nu}$. The geometry is maximally symmetric and corresponds to a conformal Poincar\'e patch of de Sitter spacetime with curvature radius $R$. Its phase space is the ordinary cotangent bundle, so the relation $(E^{2}-\vp^{2}c^{2})s^{2}=m_{0}^{2}c^{4}$ is the conformal-coordinate expression of the standard covariant mass shell. The quantum dynamics consequently follow from conventional curved-spacetime field actions and canonical equal-time structures. The geometry is a constant-curvature completion of the FL conformal factor rather than a consequence of FL kinematics, and the sections developing it record de Sitter field theory in the FL parametrization.

For the scalar field, the covariant Klein--Gordon operator contains a Hubble-type first-derivative term. Canonical rescaling eliminates this term and produces the curvature contribution $(12\xi-2)c^{2}/(R^{2}s^{2})$. For the spinor field, the tetrad and spin connection yield a conformally rescaled Dirac equation with the coefficient $\mapp=m_{0}/s$. The squared spinor operator contains the correction proportional to $\dot\mapp\gamma^{0}$, which has a direct geometric origin in the de Sitter spin connection.

The one-dimensional oscillator sectors can be solved analytically at fixed time. They are coordinate-defined models, the trap being introduced at the level of the canonical equations in the conformal chart rather than generated by a covariant trapping field, and their instantaneous spectra depend on the prescription adopted for the trapping frequency. A frequency held fixed with respect to conformal coordinate time produces an apparent spectral drift and a time-dependent coordinate localization length, whereas a frequency held fixed in the local orthonormal frame removes the leading drift from the rescaled eigenvalue. The Klein--Gordon sector is treated throughout as a second-order system, with adiabaticity controlled by the WKB ratio $\hbar|\dot E_{n}|/E_{n}^{2}$ and by basis-derivative couplings compared with the relevant level and particle--antiparticle gaps, and with the remaining mixing described by Bogoliubov coefficients or the Feshbach--Villars formulation.

The principal conclusion is that the spectral drift obtained by holding a conformal-coordinate trap frequency fixed follows from the chart and from the frequency prescription: it is removed by a change of prescription together with a rescaling. The rescaled quantity that remains is a representation-dependent eigenvalue rather than a measured local spectrum, and the $R^{-2}$ term appearing in it originates in the field redefinition $\chi=a\Phi$ rather than in the curvature. Determining which curvature effects are operationally visible requires a covariantly defined trapping field, a clock and a detector model, preferably in Fermi normal coordinates, together with dimensionless comparison standards. The present analysis provides a basis for such a calculation by specifying the field equations and oscillator models in the conformal chart.
\appendix

\section{Curvature calculation}
\label{app:curvature}

For $g_{\mu\nu}=a^{2}\eta_{\mu\nu}$ with $a=s^{-1}$, the conformal derivative is $\Ham=a'/a=-1/(Rs)$ and
\be
 \Ham'=\frac{1}{R^{2}s^{2}}=\Ham^{2}.
\label{eq:Hprime}
\ee
Substituting Eq.~(\ref{eq:christoffel}) into the definition of the Riemann tensor gives Eq.~(\ref{eq:riemann}). The Ricci scalar may also be obtained from the flat-FLRW conformal formula
\be
 \Ric=-6\frac{a''}{a^{3}}=-\frac{12}{R^{2}},
\label{eq:Rscalar-appendix}
\ee
where the sign corresponds to the convention stated in Sec.~\ref{sec:geometry}. A symbolic verification script is included in the source package.

\section{Scalar field rescaling}
\label{app:scalar}

Starting from Eq.~(\ref{eq:KG-conformal-time}) and setting $\Phi=\chi/a$,
\bea
 \Phi'&=&\frac{1}{a}\left(\chi'-\Ham\chi\right),\nonumber\\
 \Phi''&=&\frac{1}{a}\left[\chi''-2\Ham\chi'
 +(\Ham^{2}-\Ham')\chi\right].
\label{eq:Phi-derivs}
\eea
The combination $\Phi''+2\Ham\Phi'$ becomes
\be
 \frac{1}{a}\left[\chi''-\frac{a''}{a}\chi\right].
\label{eq:friction-remove}
\ee
Since $a''/a=2/(R^{2}s^{2})$, Eq.~(\ref{eq:chi-eq-x0}) follows immediately.

\section{Spin connection and conformal spinor}
\label{app:spin}

For the tetrad in Eq.~(\ref{eq:tetrad}), the nonvanishing spin-connection contribution can be represented as
\be
 \Gamma_{i}=\frac{\Ham}{2}\gamma_{i}\gamma_{0},\qquad \Gamma_{0}=0,
\label{eq:Gamma-spin}
\ee
which implies
\be
 \gamma^{\mu}(x)\Gamma_{\mu}=\frac{3\Ham}{2a}\gamma^{0}.
\label{eq:spin-contract}
\ee
Multiplying the covariant Dirac equation by $a$ gives Eq.~(\ref{eq:dirac-physical}). With $\psi=a^{-3/2}\Psi$,
\be
 \left(\partial_{t}+\frac{3}{2}\frac{\dot a}{a}\right)\psi
 =a^{-3/2}\partial_{t}\Psi,
\label{eq:spin-rescale-check}
\ee
which proves Eq.~(\ref{eq:dirac-rescaled}).

\section{Squaring the rescaled Dirac operator}
\label{app:square}

Let
\be
 D_{-}=i\hbar\gamma^{0}\partial_{t}+i\hbar c\gamma^{i}\partial_{i}-\mapp c^{2},
\qquad
 D_{+}=i\hbar\gamma^{0}\partial_{t}+i\hbar c\gamma^{i}\partial_{i}+\mapp c^{2}.
\label{eq:Dpm}
\ee
Using the Clifford algebra and allowing the derivative to act on $\mapp(t)$,
\be
 D_{+}D_{-}=-\hbar^{2}(\partial_{t}^{2}-c^{2}\nabla^{2})
 -i\hbar c^{2}\dot\mapp\gamma^{0}-\mapp^{2}c^{4}.
\label{eq:Dproduct}
\ee
Multiplication by $-1/\hbar^{2}$ gives Eq.~(\ref{eq:dirac-square}).

\end{document}